\documentclass[11pt]{article}
\usepackage{moriond,epsfig}

\def\al{\alpha}
\def\as{\alpha_{\mbox{\scriptsize s}}}

\def\ee{e^+e^-}
\def\LQCD{\Lambda_{\mbox{\scriptsize QCD}}}

\def\GeV{\mathop{\rm Ge\!V}}
\def\xB{x_{\mbox{\scriptsize B}}}

\def\gam{\gamma}

\def\out{\mbox{\scriptsize out}}

\def\cO#1{{\cal{O}}\left(#1\right)}

\def\VEV#1{\left\langle#1\right\rangle}

\def\PT{\mbox{\scriptsize PT}}
\def\NP{\mbox{\scriptsize NP}}

\def\cp{\lambda^{\NP}}

\def\ka{\kappa}

\def\cM{{\cal{M}}}

\def\cP{{\cal {P}}}

\def\Journal#1#2#3#4{{#1} {\bf #2} (#3) #4}

\def\NPB{{\em Nucl. Phys.} B}
\def\PLB{{\em Phys. Lett.}  B}
\def\PRL{{\em Phys. Rev. Lett.}}
\def\PRD{{\em Phys. Rev.} D}

\def\JHEP{{\em J. High Energy Phys.}}
\def\EPJC{{\em Eur. Phys. J.} C}
\def\EPJD{{\em Eur. Phys. J. direct} C}
\def\ibid{{\em ibid.}}

\begin{document}
\begin{flushright}
{
  BICOCCA-FT-02-13\\
  hep-ph/0207072 \\
}
\end{flushright}

\vspace*{2cm}
\title{AZIMUTHAL CORRELATION IN DIS 
\footnote{Talk presented by GM at the XXXVIIth Rencontres de Moriond
`QCD and high energy hadronic interactions', Les Arcs, France, and by
AB at the X International Workshop on Deep Inelastic Scattering
(DIS2002), Krakow, Poland.}}

\author{ANDREA BANFI, GIUSEPPE MARCHESINI AND GRAHAM SMYE}

\address{Universit\`a di Milano-Bicocca and INFN, Sezione di Milano, Italy}

\maketitle\abstracts{
We introduce a new angular correlation in DIS process and study its
differential distribution in the region in which the observable is
small. We perform a perturbative resummation at single logarithmic
accuracy and estimate leading non-perturbative power corrections.  }

\section{Introduction}
Energy-energy correlation~\cite{EEC,PP} was one of the first collinear
and infrared safe (CIS) observables studied in QCD. It involves the polar
angle between two produced hadrons in $\ee$ annihilation.

An analogous observable in a process with incoming hadrons
should involve an azimuthal angle. In DIS we then define the
azimuthal correlation~\cite{azimuth}
\begin{equation}
\label{eq:H}
H(\chi)=\sum_{a,b} \frac{p_{ta}p_{tb}}{Q^2} \delta(\chi-\chi_{ab})\>,\quad
\chi_{ab}=\pi-|\phi_{ab}|\>,
\end{equation}
being $\phi_{ab}$ the angle between $\vec p_{ta}$ and $\vec p_{tb}$,
the transverse momenta of produced hadrons $a$ and $b$ with respect
to the photon axis in the Breit frame. 
The differential distribution in $H(\chi)$ takes its first non-zero
contribution at order $\as^2$, so that its study is better performed
in DIS events with two high $p_t$ jets. Such events can be selected
for instance by constraining the two-jet resolution variable
$y_2$.\cite{DISkt}

\section{The observable at parton level}
Since the observable is linear in outgoing particle momenta, we
can replace the sum over hadrons in eq.~\ref{eq:H} with a sum over
partons. The Born process is $q \>P_1 \to P_1\> P_2$, with
$q$ the virtual boson, $P_1$ the struck parton inside the proton,
$P_2$ and $P_3$ two outgoing hard partons. Since $P_2$ and $P_3$ are
in the same plane, we have $P_{t2}=P_{t3}=P_t$ and $\chi_{23}=0$.

Beyond Born level the QCD process is $q\> p_1 \to p_2\> p_3\> k_1
\dots k_n$, with $k_i$ secondary partons. Outgoing partons $p_a$
($a=2,3$) are displaced from $P_a$ by soft recoil components, so that
$\chi_{23}$ no longer vanishes. Furthermore, one has to consider also
the correlation of secondary partons $k_i$ with $p_2$ or $p_3$.

For small $\chi$ we can consider $k_i$ soft and, to first non-trivial
order, $H(\chi)$ can be split into a hard and a soft
piece:
\begin{eqnarray}
H_h(\chi)&=&\frac{2P_t^2}{Q^2}\delta(\chi-|\phi_x|)\>,\qquad
\phi_x=\frac{\sum_i k_i^{\out}}{P_t}\>,\\
H_s(\chi)&=&\frac{2P_t^2}{Q^2} \sum_i\frac{k_{ti}}{P_t}
\left(\delta(\chi-|\bar\phi_i-\phi_x|)-
|\cos\bar\phi_i|\delta(\chi-|\phi_x|)
\right)\>.
\end{eqnarray}
The hard piece $H_h(\chi)$ depends on the total out-of-plane recoil of
$p_2$ or $p_3$, which is proportional to the sum of $k^{\out}_i$, the
out-of-event-plane\footnote{The event plane can be identified for
instance by the Breit axis and the thrust-major axis.\cite{disko}
} momenta of emitted particles. 
The soft piece
$H_s(\chi)$ represents the correlation of $k_i$ with $p_2$ and
$p_3$. The angle $\bar\phi_i$ is $\phi_{i2}$ ($\phi_{i3}$) for $k_i$
near $p_2$ ($p_3$). The subtraction term comes from the in-plane
recoil of $p_2$ and $p_3$ and ensures that the observable is CIS.

\section{Perturbative resummation \label{sec:PT}}
The first order PT contribution to the differential $H(\chi)$ distribution 
$d\Sigma/d\chi$ is given by:
\begin{equation}
\chi\frac{d\Sigma(\chi)}{d\chi}=
\frac{2\as C_T}{\pi}\ln\frac{1}{\chi}+\dots\>,\qquad C_T=2C_F+C_A\>.
\end{equation}
The presence of such a logarithm is due to an incomplete real-virtual
cancellation. Moreover, terms of the form $\as^m\ln^n\chi$ arise at
any order in PT theory. For small $\chi$ they become large and need to
be resummed to give meaning to the perturbative expansion.

Resummation of logarithmic enhanced terms can be achieved by
introducing the impact parameter $b$, the Fourier variable conjugate
to $\phi_x$. We aim at single logarithmic (SL) accuracy, i.e. having
under control all double ($\as^n\ln^{n+1} \chi$) and single ($\as^n\ln^n
\chi$) logarithms in $\ln\Sigma$. We find that only $H_h(\chi)$
contributes to SL level, while the effect of $H_s(\chi)$ is subleading 

The PT answer can then be written as a convolution of the Born matrix
element $M_0^2$ with a function which resums all double
and single logs:
\begin{equation}
\label{eq:Sigma-PT}
\frac{d\Sigma^{\PT}(\chi)}{d\chi} \sim M_0^2\otimes 
\frac{2}{\pi}
\int_0^{\infty}\!\!\! P_t db\cos(b P_t\chi)\cP(b^{-1})
e^{-R(b)}\>.
\end{equation}
In the above expression, due to coherence of QCD radiation, (virtual)
gluons with frequencies below $b^{-1}$ reconstruct the proper hard
scale for the parton density $\cP$ of the incoming parton. Gluons with
frequencies above $b^{-1}$ build up the Sudakov exponent $R(b)$, the
'so-called' radiator.  This function, the same occurring in three-jet
event shapes,\cite{disko,3jet} depends on the colour charges of the
three hard emitters and on the geometry of the hard underlying event.

The behaviour of $d\Sigma/d\chi$ near $\chi=0$
can be understood by considering the physical effects which can keep
the angle $\chi$ small. One mechanism is radiation suppression.
This is the only one relevant in most event shapes and gives rise to a
Sudakov form factor with a characteristic peak.\cite{PTstandard}
However, since our observable measures radiation only through hard
parton recoil, it may happen that $\chi$ is kept small by successive
cancellation of larger out-of-plane momenta.  It is this effect which
prevails at small $\chi$, so that $d\Sigma/d\chi$, unlike event
shapes, has no Sudakov peak, but rather approaches a constant for
$\chi \to 0$.\cite{PP}
					
\section{Non-perturbative power corrections}
The soft term $H_s$, although subleading at PT level, gives rise to the
leading NP power corrections:
\begin{equation}
\frac{d\Sigma^{\NP}(\chi)}{d\chi}\sim
M_0^2\otimes\frac{2}{\pi}\int_0^{\infty}\!\!\! P_t db \cos(bP_t\chi)
P_1(b^{-1})e^{-R(b)}B(b)\>,
\end{equation}
Extracting the term of $B(b)$ linear in $b$ we find:
\begin{equation}
\label{eq:B}
B(b) = -b\> (C_2+C_3) \int_0^{Q^2} \frac{d\ka^2}{\ka^2} 
\ka\frac{\as(\ka)}{\pi}\>.
\end{equation}
In this expression $C_2$ and $C_3$ are the colour charges of the two
outgoing partons $P_2$ and $P_3$, and $\ka$ is the (invariant)
transverse momentum of the emitted gluon with respect to each outgoing
parton.

Equation \ref{eq:B} involves the integral of the running coupling in
the infrared. Giving sense to that integral requires a genuine NP
input which can be provided for instance by the dispersive
approach.\cite{DMW} This includes the following steps:
\begin{itemize}
\item
extension of the coupling in the infrared via a dispersion relation;
\item 
promoting the gluon to be massive: this allows the gluon to
decay inclusively;
\item
taking into account non-inclusiveness of the observable by multiplying
the result by the Milan factor $\cM$.\cite{Milan}
\end{itemize}
The final result thus becomes:
\begin{equation}
B(b)=-b\>(C_2+C_3) \cp\>,\qquad \cp=\cM\frac{4}{\pi^2}\mu_I
\left(\al_0(\mu_I)+\cO{\as}\right)\>.
\end{equation}
The quantity $\al_0$ is the average of the dispersive coupling below
the (arbitrary) infrared scale $\mu_I$. This NP parameter has been
measured through the analysis of mean values and distributions of
two-jet event shapes both in $\ee$ annihilation~\cite{alfa0-ee} and
DIS.\cite{alfa0-DIS}

The NP contribution to the azimuthal correlation distribution is then:
\begin{equation}
\label{eq:Sigma-NP}
\frac{d\Sigma^{\NP}(\chi)}{d\chi}=-\cp(C_2+C_3)\VEV{b}\>,
\qquad \VEV{b}_{\chi=0}\sim\frac{1}{\LQCD}
\left(\frac{\LQCD}{Q}\right)^{\gam}\>.
\end{equation}
Here $b$ is averaged over the PT distribution in
eq.~\ref{eq:Sigma-PT}. From eq.~\ref{eq:Sigma-NP} we see that power
corrections scale like a non integer power of $1/Q$ ($\gam\simeq
0.62$).

\section{Results and conclusions}
\begin{figure}[htp]
\begin{minipage}{0.5\textwidth}
\epsfig{file=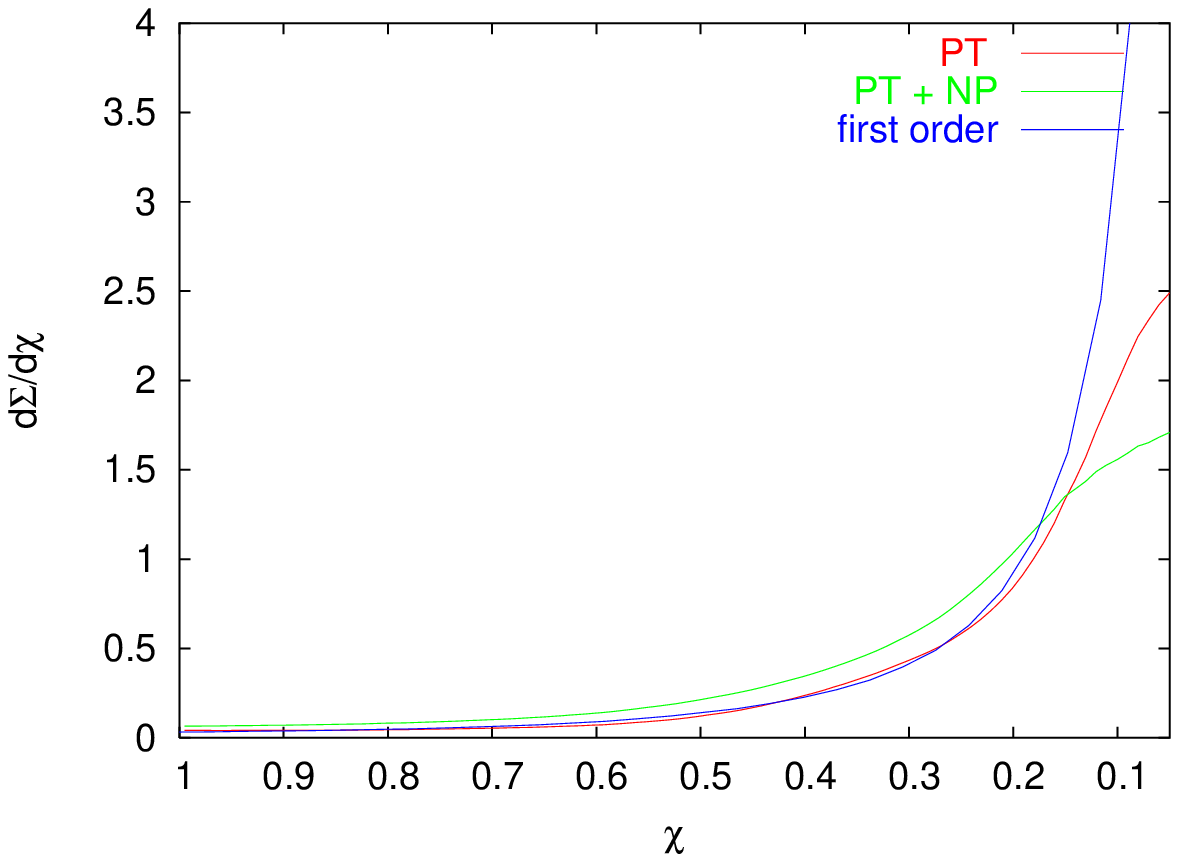, width=\textwidth}
\end{minipage}
\begin{minipage}{0.4\textwidth}
\epsfig{file=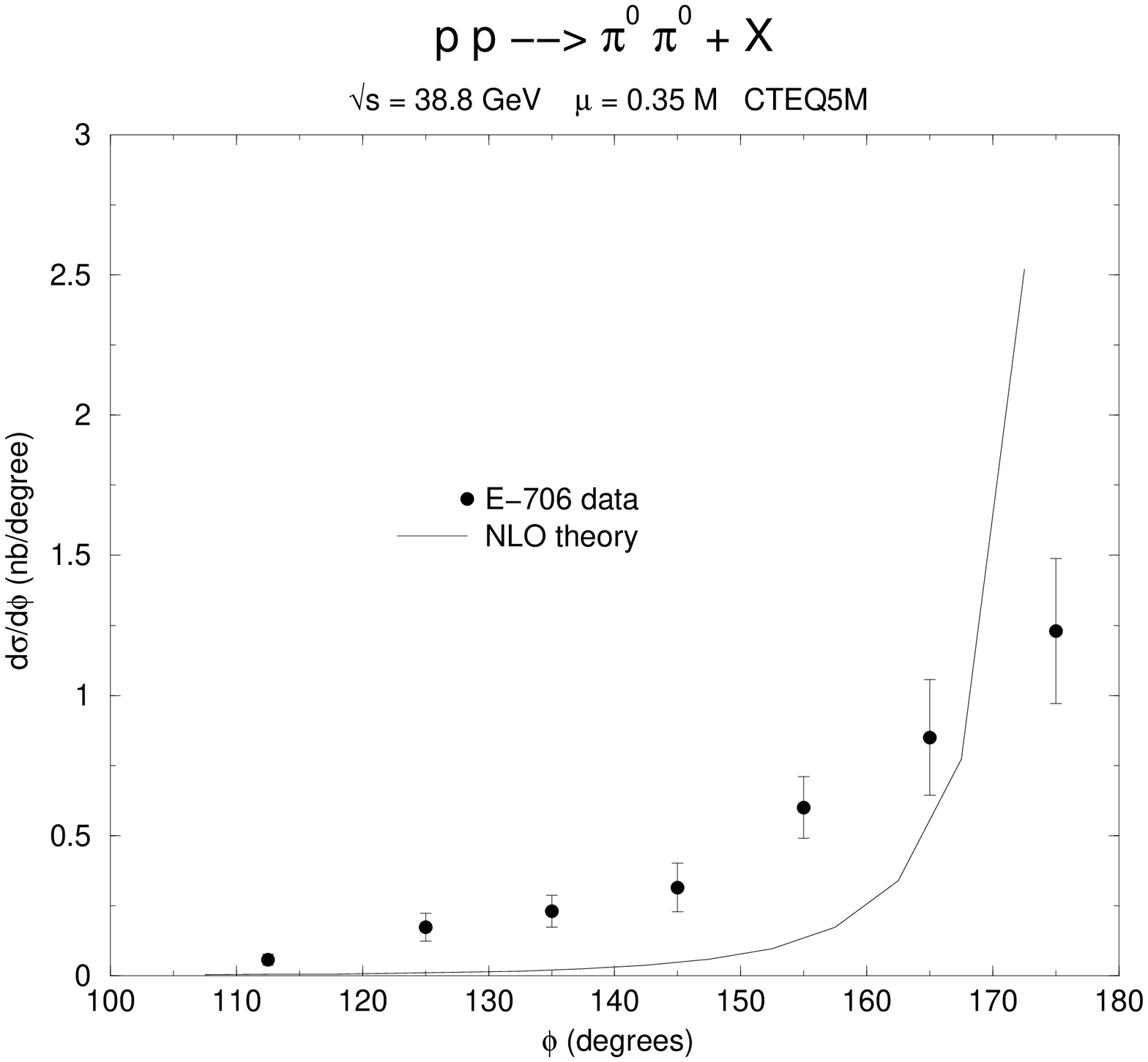, width=0.9\textwidth, angle=270}
\end{minipage}
\caption{Qualitative comparison between azimuthal correlation in DIS (a) and 
$\pi^0 \pi^0$ azimuthal correlation in hadron-hadron collisions (b).
\label{fig:plot}}
\end{figure}

In figure~\ref{fig:plot}a we show the behaviour of azimuthal
correlation at HERA energies for $Q^2=900\GeV^2$, $\xB=0.1$ and
$1.0<y_2<2.5$. We see that the distribution goes to constant for $\chi
\ll 1$. The effect of power corrections is to further deplete the
distribution by an amount proportional to $\VEV{b}$.  In
figure~\ref{fig:plot}b we report also the $\pi^0 \pi^0$ azimuthal
correlation taken from E706 data.\cite{Owens} We observe that also in
this case the distribution flattens to a constant value thus suggesting
the mechanism of cancellation of out-of-plane momenta discussed in
section~\ref{sec:PT}.
In conclusion, we have now a new observable that can be used not only
to provide a further measurement of $\as$ and to constrain the parton
densities, but also to investigate the nature of hadronisation effects
in hard QCD processes.

\section*{Acknowledgments}
We are grateful to Yuri Dokshitzer for helpful discussions and to
Gavin Salam and Giulia Zanderighi for useful comments and support in
the numerical analysis.

\end{document}